\documentclass[aps,preprint,amsmath,showpacs]{revtex4}
\usepackage{graphicx}
\usepackage[latin1]{inputenc}

\begin{document}

\title{Quantum linear mutual information and classical correlations in globally pure
bipartite systems}

\author{R.M. Angelo\footnote[1]{Present address: Instituto de F\'{\i}sica,
Universidade de S\~ao Paulo, C.P. 66318, 05315-970, S\~ao Paulo, SP, Brazil.},
S. A. Vitiello, M.A.M. de Aguiar and K. Furuya}

\affiliation{Instituto de F\'{\i}sica `Gleb Wataghin', Universidade
Estadual de Campinas, C.P. 6165, \\13083-970, Campinas, SP, Brazil}

\begin{abstract}

We investigate the correlations of initially separable probability
distributions in a globally pure bipartite system with two degrees
of freedom for classical and quantum systems. A classical version
of the quantum linear mutual information is introduced and the two
quantities are compared for a system of oscillators coupled with
both linear and non-linear interactions. The classical
correlations help to understand how much of the quantum loss of
purity are due to intrinsic quantum effects and how much is
related to the probabilistic character of the initial states, a
characteristic shared by both the classical and quantum pictures.
Our examples show that, for initially localized Gaussian states,
the classical statistical mutual linear entropy follows its
quantum counterpart for short times. For non-Gaussian states the
behavior of the classical and quantum measures of information are
still qualitatively similar, although the fingerprints of the
non-classical nature of the initial state can be observed in their
different amplitudes of oscillation.


\end{abstract}

\maketitle
%===========================================================
%===========================================================
\section{Introduction}

Entanglement has been a focus of intense investigation in the recent
years due to its relevance in quantum computation and quantum
information \cite{1,2,vedral,plenio,vidal,woot}.  From a fundamental
point of view, entanglement is considered ``the characteristic trait
of quantum mechanics, the one that enforces its entire departure from
classical lines of thought''
(Schr\"odinger)\cite{schrodinger35}. Entanglement reflects quantum
correlations in the Hilbert space, where state vectors are associated
with probability distributions. In classical mechanics, on the other
hand, states are described by points in phase space, and no intrinsic
probabilistic character exists.

Probability distributions can be introduced in classical mechanics
with the Liouville formalism, and statistical averages on
ensembles can be calculated. Initially independent probability
distributions can become correlated when evolved by a classical
Hamiltonian with interaction terms. These correlations represent
the dynamical emergence of conditional probabilities in the
classical statistical world.

In this paper we consider the time evolution of a bipartite system
whose Hilbert space is the direct product of two subspaces.
Initial states that are the tensor product of kets in each
subspace, usually evolve to non-product states, generating
coherences and entangling the two subsystems. A measure of the
non-separability between them is provided either by the Von
Neumann or the linear entropies. In the case of a globally pure
bipartite system we argue that the latter is a convenient quantum
measure of non-separability. Next, as the main point of our work,
we consider the time evolution of probability densities in the
phase space of the corresponding bipartite {\em classical}
analogue. Inspired on the quantum problem we propose a measure of
the classical correlations generated by the statistical ensembles
of classical trajectories. We want to measure the non-separability
of the classical probability distribution evolving in time via
Liouville equations. Moreover we want to compare behavior of this
classical measure with the loss of purity of the quantum system,
particularly at short times. This comparison helps to identify the
time at which intrinsic quantum effects, such as interferences,
begin to be important.

In our approach we first define a classical statistical quantity
and compare its behavior with the quantum linear entropy (QLE).
Due to its formal similarity with the quantum linear entropy, we
call it {\it classical statistical linear entropy} (CSLE).  A
similar classical quantity has recently been used in the study of
quantum-classical correspondence of intrinsic decoherence
\cite{gong03}. In spite of the direct correspondence between the
QLE and the CSLE, the latter may not be symmetric between the two
subsystems that make up the global bipartite system, i.e., the
CSLE of each subspace does not contain in general the same
information. Also, the CSLE can assume negative values and is
restricted to initially separable distributions, not being a good
measure in the case of distributions that start off non-separable.
We therefore define a more general classical measure, the {\it
classical statistical linear mutual information} (CSLMI), based on
the concept of {\it quantum mutual information} (QMI)
\cite{vedral97}, that is always symmetric and positive, and thus,
best suited to measure the classical separability.  We show that
for two bi-linearly coupled harmonic oscillators with Gaussian
initial distributions, the classical and quantum results coincide.
For non-linear coupling the classical and quantum entropies remain
close for short times for both regular and chaotic regimes.

The outline of this work is as follows. In Section II.A we give
some general definitions and show that the quantum linear entropy
can be expressed in terms of the non-diagonal terms of the density
operator, providing a good measure of the entanglement between the
subsystems. In Section II.B we present our definitions of the
classical statistical linear entropy with some considerations
about the initial probability distributions. Next we introduce the
quantum linear mutual information and its classical statistical
analogue. This allows us to treat cases where the two subsystems
have different types of initial distributions in phase space.
Section III is reserved to the comparison between the QLMI and the
CSLMI for a system of two oscillators coupled with the following
types of interactions: (1) bi-linear; (2) non-linear ; (3)
bilinear within the {\em rotating wave approximation} (RWA). As
initial states we consider both Gaussian and non-Gaussian
distributions. Finally in Section IV we present some conclusions
and discuss the adequacy and limitations of the present approach.

%===========================================================
%===========================================================
\section{Theory}

%=======================================================
\subsection{Quantum Linear Entropy}

The density operator corresponding to an initially  (normalized) pure
state $|\psi_0 \rangle$ is
\begin{equation}
\rho(t) = U(t) \, |\psi_0 \rangle \langle \psi_0| \, U^{\dagger}(t),
\end{equation}
where $U(t)=e^{-iHt/\hbar}$ is the evolution operator and $H$ is the
Hamiltonian. For a globally pure state, there are constraints
connecting the diagonal and off-diagonal matrix elements of the density
operator. This follows from the property $Tr\rho^2 = Tr\rho =1$, which
can be written explicitly as
%----------------------
\begin{eqnarray}
\sum_i\rho_{ii}²+\sum_i\sum\limits_{k\neq i}|\rho_{ik}|^2=\sum_i
\rho_{ii} = 1 \;,
\label{correlacoes}
\end{eqnarray}
%----------------------
where we have separated the diagonal and off-diagonal terms in the
left hand side. From this relation we define
%----------------------
\begin{eqnarray}
d (\rho)\equiv\sum_i\sum\limits_{k\neq i}|\rho_{ik}|²=
1-\sum_i\rho_{ii}² \;.
\label{emaranhamento}
\end{eqnarray}
%----------------------

Consider now a bipartite system whose Hilbert space is the direct
product of two sub-spaces: $\mathcal{E} = \mathcal{E}_1 \otimes
\mathcal{E}_2$. A measure of non-separability between these subsystems
for a globally pure state is provided either by the Von Neumann or by
the linear entropies \cite{vedral97,manfredi}. The partial trace of
$\rho$ on system $2$ defines the operator
\begin{equation}
\label{rho1q}
\rho_1(t) = tr_2(\rho(t)) \;.
\end{equation}
The corresponding QLE is given by
\begin{equation}
\label{d1}
S_1(t)  = 1 -  tr_1(\rho_1^2(t)) \;.
\end{equation}
The operator $\rho_2(t)$ and the QLE $S_2(t)$ are similarly defined.

For $H = H_1\otimes\mathbf{1}_2 + \mathbf{1}_1\otimes H_2$ both
$\rho_1(t)$ and $\rho_2(t)$ are projectors and $S_1(t)=S_2(t)=0$.
If couplings between the sub-spaces exist, they force the global
state to evolve to non-product states, entangling the subsystems and
producing non-zero values for $S_1(t)$ and $S_2(t)$.

In fact, for a globally pure bipartite system
$S_1=S_2=d (\rho)$. To see this we write
%----------------------
\begin{eqnarray}
|\psi\rangle=\sum_i \lambda_i |a_i\rangle|b_i\rangle,
\end{eqnarray}
%----------------------
where $\{|a_i\rangle\}$ and $\{|b_i\rangle\}$ are orthonormal basis
of subsystems $1$ and $2$ respectively, and  $\{\lambda_i\}$ is the
Schmidt spectrum \cite{schmidt07}. Then,
%----------------------
\begin{eqnarray}
\rho=
\sum_{i,j}\rho_{ij}|a_i\rangle |b_i\rangle\langle a_j| \langle b_j|,
\end{eqnarray}
%----------------------
with $\rho_{ij}=\lambda_i \lambda_j^*$. The reduced density of system
$1$ becomes
%----------------------
\begin{eqnarray}
\rho_1&=&\sum_i \rho_{ii}|a_i\rangle\langle a_i|,
\end{eqnarray}
%----------------------
and similarly for system $2$. The subsystem linear entropy becomes
trivial in this basis:
%----------------------
\begin{eqnarray}
S_1=1-Tr_{1} (\rho_{1}^2)= 1-\sum_i \rho_{ii}^2
\label{delta}
\end{eqnarray}
%----------------------
with a similar result for $S_2$.

Comparing this result with Eq.(\ref{emaranhamento}) we see that
%----------------------
\begin{eqnarray}
S_1 = S_2 = \sum_i\sum\limits_{k\neq i}|\rho_{ik}|^2=d (\rho).
\label{s1s2}
\end{eqnarray}
%----------------------
Therefore, as long as any globally pure state of the system can be
decomposed in the Schmidt basis, we can conclude that the linear
entropy indeed measures coherences between the
subsystems. Furthermore, because of the total conservation
of coherences (\ref{correlacoes}), the linear entropy
seems to be the most natural measure in the present situation.

%===========================================================
\subsection{Classical Systems}

A measure of non-separability for classical systems can be defined
only at a statistical level by considering ensembles of initial
conditions \cite{3,4,5}. Following Wehrl \cite{wehrl} we define a
quantity that we call {\it classical statistical linear entropy}.

Consider a system with two degrees of freedom described by the classical
Hamiltonian function $\cal{H}$. Consider also several copies of this
system with initial conditions distributed according to
the ensemble probability distribution
$P(x,t=0)$,
where $x \equiv (q_1,p_1,q_2,p_2)$. The classical time
evolution of $P(x,0)$ is obtained via Liouville's equation
\begin{equation}
\frac{\partial P}{\partial t} = \{\mathcal{H},P\} ,
\label{Lio}
\end{equation}
whose solution is
\begin{equation}
\label{timeevol}
P(x,t) = P(\phi_t^{-1}(x),0),
\end{equation}
where $\phi_t^{-1}(x) \equiv x_0$ is the initial condition that
propagates to $x$ in the time $t$ and $\phi_t$ is the phase space
flux, so that $x=\phi_t(x_0)$.  In words, the numerical value of
the probability $P$ at point $x$ and time $t$ has the same
numerical value of the probability at point $x_0$ of the {\em
initial distribution}. This value is carried over to $x$ in the
time $t$. Liouville's theorem guarantees the conservation of
probability at all times. For integrable systems $\phi_t(x_0)$
might be determined analytically, otherwise numerical calculations
can be performed. Notice that $P(x,t)$ is itself a constant of
motion, but this is not enough to guarantee the integrability of
the system, since it is not generally in involution with the
Hamiltonian.

The marginal probability distribution
\begin{equation}
\label{p1t}
P_1(q_1,p_1,t) = \int dq_2 dp_2 P(q_1,p_1,q_2,p_2,t)
\end{equation}
allows us to define the {\em classical statistical linear entropy}
\begin{equation}
\label{deltaclas}
S_1^{cl}(t) = 1 - \frac{ \int dq_1 dp_1 P_1^2(q_1,p_1,t)}
{ \int dq_1 dp_1 P_1^2(q_1,p_1,0)} \;.
\end{equation}
The normalization is necessary for dimensional reasons and to
guarantee that $S_1^{cl}(0)=0$.

The initial classical distribution corresponding to a given quantum
pure state $\rho(0)$ is chosen as the coherent states phase space
projection
\begin{equation}
\label{classicalp}
P(x,0) =\frac{1}{N}
    \langle\alpha_1,\alpha_2|\rho(0)|\alpha_2,\alpha_1\rangle,
\end{equation}
where $N$ is a normalization constant and $\alpha_i=(q_i+\imath
p_i)/\sqrt{2\hbar}$ is the usual complex parametrization of the
coherent states. Eq. (\ref{classicalp}) is the normalized Husimi
distribution, which is positive definite by construction. It is known
that the Husimi distribution does not reproduce the correct marginal
probabilities. However, the constraint of a positive probability
distribution excludes the Wigner function  \cite{6b} as a classical
initial distribution.

Eqs.(\ref{emaranhamento}) and (\ref{s1s2}) show that the quantum
linear entropies $S_1$ and $S_2$ can be obtained only from the
diagonal elements of the global density matrix. Diagonal elements, on
the other hand, have classical analogues. Therefore it makes sense to
compare the dynamics of $S_1$ and $S_2$, which may be written either
in terms of only off-diagonal or only diagonal elements, with the
classical correlations.

An important difference between the classical and quantum linear
entropies can be made explicit in the coherent state basis. For
one dimensional systems we obtain
\begin{equation}
S(t)=1-\int \frac{d²\alpha}{\pi} \Big[\langle\alpha|U(t)\rho²(0)U^{\dag}(t)
|\alpha\rangle\Big].
\end{equation}
On the other hand, defining the Liouvillian operator $\cal{L}$ such
that $\{\mathcal{H},P\}=\mathcal{L}P$, Eq.(\ref{Lio}) can be formally
integrated and the CSLE can be written as
\begin{equation}
\label{classicals}
S^{cl}(t)=1-\int \frac{d²\alpha}{\pi}
         \left[\frac{\left(e^{\mathcal{L}t}
        \langle\alpha|\rho(0)|\alpha\rangle\right)^2}{M}\right],
\end{equation}
where $M=N²/\pi$. In particular, at $t=0$ the integrand of the
classical entropy depends on $\langle \alpha
|\rho(0)|\alpha\rangle^2$ instead of
$\langle\alpha|\rho^2(0)|\alpha\rangle$. Another important
difference is of course in the dynamics: the quantum evolution is
determined by non-commuting operators, which brings a number of
corrections to the classical formalism.  For more degrees of
freedom, although the time evolution cannot be expressed in such a
simple way, the two differences pointed out above remain true.

%===========================================================
\subsection{Quantum and Classical Linear Mutual Information}

Although the CSLE seems to be the natural classical analague of
the quantum linear entropy, it is not symmetric between the
subsystems. Indeed, if the initial probability distributions of
each subsystem are not equal (for example a Gaussian distribution
for one subsystem and a Poisonian distribution for the other) we
find that $S_1^{cl}(t) \neq S_2^{cl}(t)$. The quantum linear
entropy, on the other hand, always satisfies $S_1(t) = S_2(t)$.
Another drawback of the definition Eq.(\ref{deltaclas}) is that it
gives $S_1^{cl}(0)=S_2^{cl}(0)=0$ for any initial classical
distribution, correlated or not.  Thus, it is interesting to
define a more general quantity that avoids these difficulties and
that takes into account the contributions of both subsystems
entropies symmetrically. At the quantum level we define the {\sl
quantum linear mutual information} (QLMI) that depends on the QLE
as $I\equiv S(\rho_1\otimes\rho_2)-S(\rho)$. This is based on the
{\sl Von Neumann mutual information} \cite{vedral97,11},
$I_{VN}=S_{VN}(\rho_1 \otimes\rho_2)-S_{VN}(\rho)$, where $S_{VN}$
is the Von Neumann entropy. By using Eq.(\ref{d1}) we can write
the QLMI as (notice that the linear entropy is not additive
\cite{manfredi})
\begin{equation}
I(t)=S_1(t)+S_2(t)-S_1(t)\,S_2(t)-S(t),
\label{iq}
\end{equation}
where $S_{i}$ and $S$ are the subsystem and the global linear
entropies respectively.  For pure initial states $S(t)=0$ and
$I(t)=S_1+S_2 -S_1 S_2$.

We also define a quantity that we call {\sl classical statistical
linear mutual information} (CSLMI) as
\begin{equation}
I^{cl}(t)=S_1^{cl}(t)+S_2^{cl}(t)-S_1^{cl}(t)\,
S_2^{cl}(t)-S^{cl}(t).
\label{icl}
\end{equation}
Once again $S^{cl}(t)=0$ for pure states. Note that the above
definition can also be written in form $I^{cl}=S^{cl}(P_1 \,
P_2)-S^{cl}(P)$ where $S^{cl}(P)$ is given by Eq.(\ref{deltaclas})
with $P_1$ replaced by $P$.

We emphasize that the quantum and classical linear mutual information
defined above are also non-separability measures, like the QLE and
CSLE.  This can be made explicit by re-writing Eqs.(\ref{iq}) and
(\ref{icl}) as
\begin{eqnarray}
I(t)&=&tr[\rho²(t)-\rho_1²(t)\otimes\rho_2²(t)]\nonumber \\
    &=& 1- tr[\rho_1²(t)\otimes\rho_2²(t)],\\ \nonumber \\
I^{cl}(t)&=&\frac{\int dx [P²(t)-P_1²(t)P_2²(t)]}
{\int dx P²(0)}\nonumber \\
    &=& 1-\frac{\int dx P_1²(t)P_2²(t)}
{\int dx P²(0)} .
\label{iqic}
\end{eqnarray}
The last equality is true since ${\int dx P^2(0)} ={\int dx P^2(t)}$.
When the two subsystems have the same type of initial
distributions, both $I$ and $I^{cl}$ have the same contents of their
respective linear entropies. Moreover, since $I^{cl}$ is symmetric
with respect to the two subsystems, it does not present difficulties
when the initial distributions are different.

In the next section we compare $I(t)$ with $I^{cl}(t)$ for three
different cases.

%===========================================================
%===========================================================
\section{Results}
In order to show that our definition of a CSLMI is physically sensible,
we consider the following classical Hamiltonian with two degrees of
freedom
\begin{equation}
{\cal H}(q_1,p_1,q_2,p_2)=  {\cal H}_1 +  {\cal H}_2  + {\cal H}_I \; ,
\label{Hcl}
\end{equation}
where ${\cal H}_i = \frac{1}{2}(p_i^2+\omega_i^2 q_i^2)$ are harmonic
oscillators and ${\cal H}_I$ is an interaction term. In what follows
we shall calculate both CSLMI and QLMI for three different couplings: a
simple bilinear, a non-integrable and a rotating-wave approximation
(RWA). In the first and
last cases the calculations can be performed analytically.

%%%%%%%%%%%%%%%%%%%%%%%%%%%%%%%%%%%%%%%%%%%%%%%%%%%%
1) {\em Bilinear Coupling  ${\cal H}_I(q_1,p_1,q_2,p_2)= \lambda
q_1 q_2$} \cite{7}\\
In this case, the classical equations of motion can be easily
integrated and we find
\begin{eqnarray}
\label{solosc}
q_{1}(t) = q_1 C_+(t) + q_2 C_-(t) + p_1 S_+(t) + p_2 S_-(t), \\
p_{1}(t) = p_1 C_+(t) + p_2 C_-(t) - q_1 S_+(t) - q_2 S_-(t), \nonumber
\end{eqnarray}
and similar expressions for $q_{2}(t)$ and $p_{2}(t)$. $C_{\pm}$ and
$S_{\pm}$ are quasi-periodic functions of time
\begin{eqnarray}
\label{cssn}
C_{\pm}(t) = \frac{1}{2} \left(\cos \Omega_x t \pm  \cos \Omega_yt\right),
\nonumber \\
S_{\pm}(t) = -\frac{1}{2}\left({\Omega_x} \sin\Omega_x t \pm {\Omega_y}
\sin \Omega_yt \right),
\end{eqnarray}
where $\Omega_x=\sqrt{\omega^2+\lambda}$ and
$\Omega_y=\sqrt{\omega^2-\lambda}$.

As initial phase space distribution we choose a Gaussian centered at
$x_c=(q_{1c},p_{1c},q_{2c},p_{2c})$,
\begin{eqnarray}
P(x,0)=\frac{1}{4\pi²\hbar²}\exp\left\{-\frac{(x-x_c)^T(x-x_c)}{2\hbar}
  \right\}.
\label{rho0}
\end{eqnarray}
This is the classical density as given by Eq.(\ref{classicalp}),
corresponding to a coherent state initial wave-function.

The probability distribution at time $t$ is obtained using
Eqs. (\ref{timeevol}) and (\ref{solosc}). The result can be written
in the form
\begin{eqnarray}
P(x,t)= \frac{1}{4\pi^2\hbar^2} e^{ -x^T {\bf A} x + 2 B x + C},
\label{probosc}
\end{eqnarray}
where the matrix ${\bf A}$ and the vector $B$ are functions of
$C_{\pm}$, $S_{\pm}$ and $\hbar$. Replacing Eq.(\ref{probosc}) into
(\ref{icl}) we get
\begin{equation}
\label{deltaclassosc}
I^{cl}(t) = 1 - \frac{1}{64\hbar^6} \,
   \frac{1}{\det{(\beta)}^2 \det{(\alpha-\gamma \beta^{-1} \gamma^T)}},
\end{equation}
where $\alpha$, $\beta$ and $\gamma$ are $2$ x $2$ matrices
whose elements are combinations of $C_{\pm}$, $S_{\pm}$ and
$T_{\pm}$, with
\begin{eqnarray}
T_{\pm}(t) = -\left(\frac{\sin \Omega_x t}{\Omega_x} \pm
              \frac{\sin \Omega_y t}{\Omega_y} \right) \nonumber \;.
\end{eqnarray}
It is interesting to note that all the dependence on the center
position of the initial distribution has gone.

The reduced QLMI can also be analytically computed. Here, we simply write
down the result:
\begin{equation}
\label{deltaquantosc}
I(t) = 1 - \left[ \frac{4
\hbar^2(1-\lambda^2)}{|D_x(t)| |D_y(t)| \,
   \sqrt{\det{(O)}}} \right]^4.
\end{equation}
The coefficients $D_i(t)$ are periodic functions of frequency
$\Omega_i$. $O$ is an $8\times 8$ quasi-periodic matrix depending
on both $\Omega_1$ and $\Omega_2$. Expressions
(\ref{deltaclassosc}) and (\ref{deltaquantosc}) are actually
identical, which shows that our definition of $I^{cl}$ and the
choice of its normalization are both appropriate. Fig. 1 displays
an example for $\lambda = 0.9$ that shows $I_1(t) = I_1^{cl}(t)$.

%%%%%%%%%%%%%%%%%%%%%%%%%%%%%%%%%%%%%%%%%%%%%%%%%%%%
2) {\em Nonlinear Coupling} \\
The exact coincidence between the classical statistical and quantum
linear mutual informations just presented, certainly has to do with both the
quadratic nature of the Hamiltonian and the Gaussian initial distributions.
To study the role of non-linearities we consider the interaction Hamiltonian
\begin{equation}
{\cal H}_I(q_1,p_1,q_2,p_2) = - q_1p_1p_2+\frac{1}{2} {q_1}^2 {q_2}^2.
\label{Nel}
\end{equation}
The total Hamiltonian ${\cal H}$ is a canonically transformed version
of a well studied system known as the Nelson potential
\cite{8}. Fig. \ref{fig2} shows a typical mixed Poincar\'e section
for parameter values $\omega_1=\sqrt{0.1}$, $\omega_2=\sqrt{2}$ and
energy $E=0.05$.

The calculation of $I_1^{cl}(t)$ (Eq.(\ref{deltaclas})) now has to
be performed numerically. We use the same initial distribution,
Eq.(\ref{rho0}). $x_0=\phi^{-1}_t(x)$ was calculated using
standard Runge-Kutta routines. The integrations in Eqs.(\ref{p1t})
and (\ref{deltaclas}) was done both by Monte Carlo and by direct
trapezoidal techniques; the results of the two methods agree in
the time intervals we have considered.  In Fig.(\ref{fig3}a) we
show $I(t)$ and the corresponding $I^{cl}(t)$. The initial density
matrix is the direct product of two coherent states, and the
initial classical probability distribution is that given by
(\ref{classicalp}). The center of the coherent states is in the
{\sl chaotic} region of the corresponding Poincar\'e section. The
resemblance between the QLMI and CSLMI is quite good even after
two oscillations. Fig(\ref{fig3}b) shows a similar calculation
with the coherent states centered at the {\sl regular} region.
Once again the classical and quantum results coincide for short
times. Surprisingly, the two entropy-like quantities agree for
{\sl longer times} in the chaotic case. For both the regular and
chaotic cases, the classical and quantum calculations agree very
well for short times, although the classical mutual information is
systematically larger than its quantum counterpart for times
larger than about $1$. Also, the linear mutual information grows
faster for the chaotic case than for the regular case, in
accordance with similar previous results for the linear entropy
\cite{kyoko}. Finally we note that the classical linear entropies
Eq.(\ref{deltaclas}) may become negative. Similar behaviors were
reported in refs. \cite{wehrl, manfredi}. The linear mutual
information, on the other hand, is always positive and better
suited for measuring the classical loss of separability.

%%%%%%%%%%%%%%%%%%%%%%%%%%%%%%%%%%%%%%%%%%%%%%%%%%%%

3) {\em RWA Coupling:} ${\cal H}_I(q_1,p_1,q_2,p_2)= \lambda (q_1
q_2+p_1 p_2)$\\
This is the classical version of a rotating-wave approximation of
the interaction Hamiltonian treated in the example (1).  In this case,
Gaussian distributions in each subspace evolve coherently: $I_1(t)=
I_1^{cl}(t)=0$. However, this is a very particular case of a preferred
basis state \cite{6a,9}, and the same kind of coherent evolution is not
expected for more general initial states. For instance, in the case of
{\bf  Fock states} ($|\psi_0 \rangle=|1\rangle\otimes|1\rangle$) the
classical phase space distribution, given by the Husimi distribution, is
\begin{equation}
P(x,0) =\frac{e^{-\frac{(x-x_c)^T(x-x_c)}{2\hbar}}}{4\pi²\hbar²}
\prod\limits_{k=1}^2\left(\frac{q_k²+p_k²}{2\hbar} \right).
\end{equation}
In the analytical calculation of the CSLMI we use the
super-operator method \cite{10} modified to conform with classical
Poisson brackets computations. We find
\begin{equation}
I(t)=8u(t)[2-8 u(t)]
\end{equation}
and
\begin{equation}
I^{cl}(t)=u(t)[2- u(t)],
\end{equation}
where
\begin{equation}
u(t)=\displaystyle{\frac{\sin^2 (2\lambda t)}{32}
   \left[5+3 \cos{(4 \lambda t)}  \right]} = S_1^{cl}(t) = S_2^{cl}(t)\;.
\end{equation}
The quantum and the classical statistical linear entropies have
similar qualitative behaviors, but the amplitude of their
oscillations are markedly different. They exhibit the same
purification period $\tau_0=\frac{\pi}{2 \lambda}$.

Finally, we consider one of the subsystems in a coherent state and
the other in the number state ($|\psi_0 \rangle=|\alpha_1\rangle
\otimes|1\rangle$).  In this case, for $\alpha_1=0$, we obtain
\begin{equation}
  \begin{array}{ll}
 I(t) & =\displaystyle{\frac{\sin^2 (2\lambda t)}{8}
   \left[7 + \cos{(4 \lambda t)}  \right]} \\ \\
 I^{cl}(t) & =\displaystyle{\frac{\sin^2 (2\lambda t)}{64}
   \left[15 + \cos{(4 \lambda t)}  \right]} \;.
   \end{array}
\end{equation}
These quantities are plotted in Fig. (\ref{fig4}).  The
qualitative agreement of the oscillations is remarkable, though
the amplitudes are quite different. If the classical information
is re-scaled so that its maximum at $\pi/4$ coincides with the
quantum plot, the two curves become very similar. This might seem
a consequence of the normalization introduced in $I^{cl}$ and not
present in $I$ (see Eq.(\ref{iqic})). Unfortunately this is not
so: the classical normalization is indeed necessary (at least for
dimensional reasons) and choosing it in a case by case basis does
not seem to bring any important physical information. Our
interpretation of the differences in the amplitudes is that they
reflect the non-classical character of the initial phase space
distribution.

%=============================================
\section{Conclusion}

In this work we have defined the {\sl classical statistical linear
mutual information}, a tool that quantifies the non-separability
of classical statistical distributions representing pure states of
a bipartite Hamiltonian system. The comparison of the quantum and
classical linear mutual information provides a measure of how much
of the quantum loss of purity are due to intrinsic quantum effects
and how much is related only to the probabilistic character of the
initial distributions. We computed the classical and quantum
mutual information for a system of two oscillators subjected to
different types of coupling. We found that the two measures follow
each other closely in the case of initially separable Gaussian
states. For the case of linear coupling the classical and quantum
mutual information are identical, revealing the classical nature
of the system and the coherent evolution of the Gaussian
wave-packets. For non-linear couplings the classical mutual
information follows the quantum one for short times. This is
follows from the fact that the short time quantum evolution can be
formulated in terms of Liouville formalism
\cite{ballentine,angelo}. This property is desirable for a measure
of classical correlations in view of Ehrenfest's theorem
\cite{ehren}, and confirms that our definition is appropriate. For
longer times the folding of the wave-packets certainly introduces
self-interferences that have no classical counterpart. When
quantum interferences become substantial \cite{rf} the classical
and quantum mutual information become significantly different. For
open systems, where interferences are eliminated by the coupling
with an external environment, we conjecture that the classical and
quantum mutual information are going to coincide for much longer
times.

We have also investigated the role of non-Gaussian types of
initial distributions in the classical separability. We have shown
that also in this case the CSLMI $I_{cl}(t)$ is a meaningful
quantity to measure the classical non-separability. Our example of
linearly coupled harmonic oscillators (RWA) shows that the time
evolution of the classical mutual information is qualitatively
similar to that of its quantum counterpart. The amplitude of the
classical oscillations, however, is markedly different from the
quantum ones, reflecting the non-classical nature of the initial
state.

\begin{center}
{\bf Acknowledgments}
\end{center}

We acknowledge Conselho Nacional de Desenvolvimento
Cient\'{\i}fico e Tecnol\'ogico (CNPq) and Funda\c{c}\~ao de
Amparo a Pesquisa de S\~ao Paulo (FAPESP)(Contract No.02/10442-6)
for financial support.

\newpage
%%%%%%%%%%%%%%%%%%%%%%%%%%%%%%%%%%%%%%%%%%%%%%%%%%%%
%%%%%%%%%%%%% FIGURES %%%%%%%%%%%%%%%%%%%%%%%%%%%%%%
%%%%%%%%%%%%%%%%%%%%%%%%%%%%%%%%%%%%%%%%%%%%%%%%%%%%

\begin{figure}[ht]
\centerline{\includegraphics[scale=0.4,angle=-90]{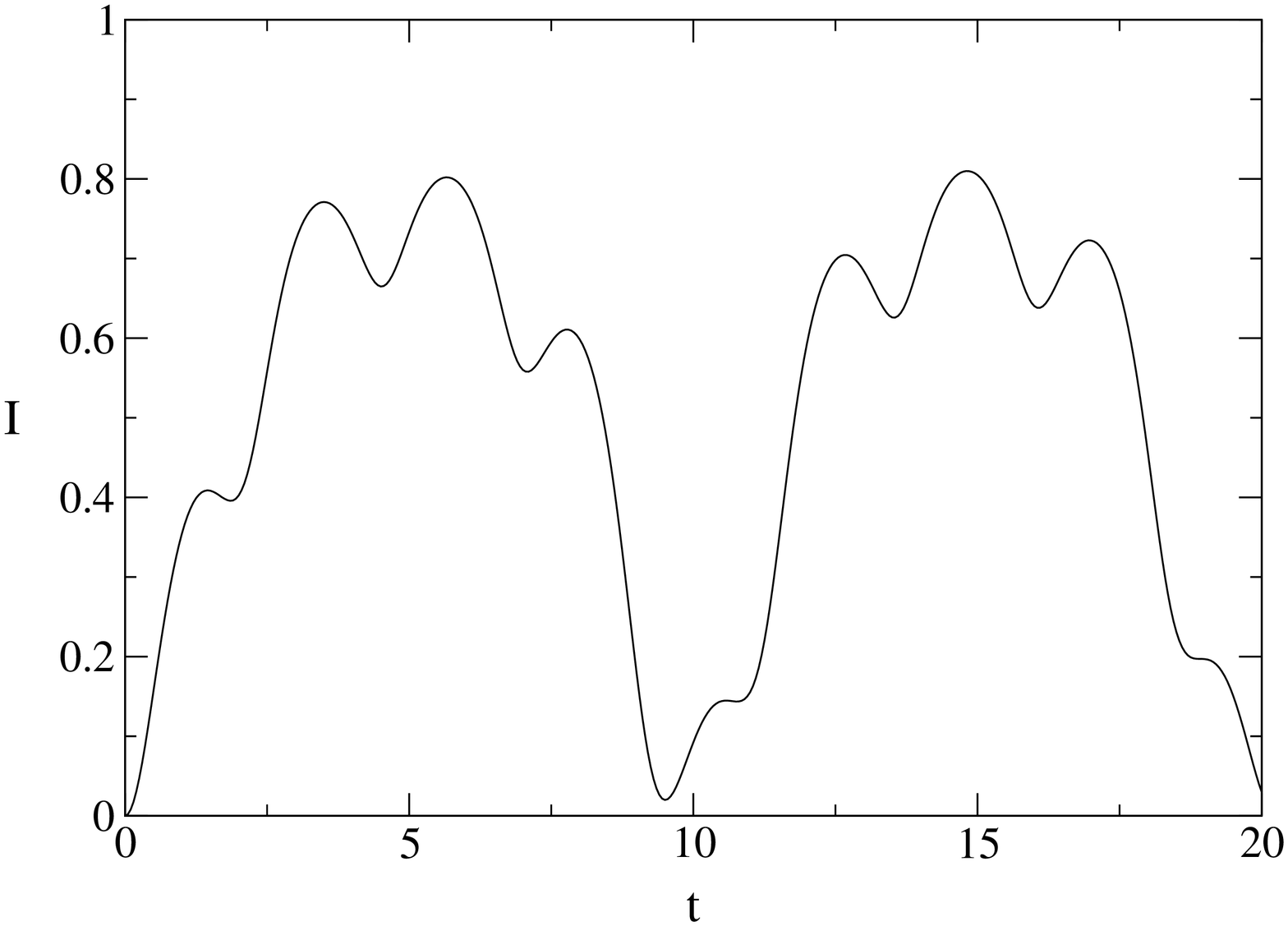}}
\caption{QLMI $I(t)$ and CSLMI $I_{cl}(t)$ as a function of time
(arbitrary units) for the linear coupling and initial Gaussian
distribution. The two curves are identical.}
\label{fig1}
\end{figure}

\begin{figure}[ht]
\vspace{0.3cm}
\centerline{\includegraphics[scale=0.4]{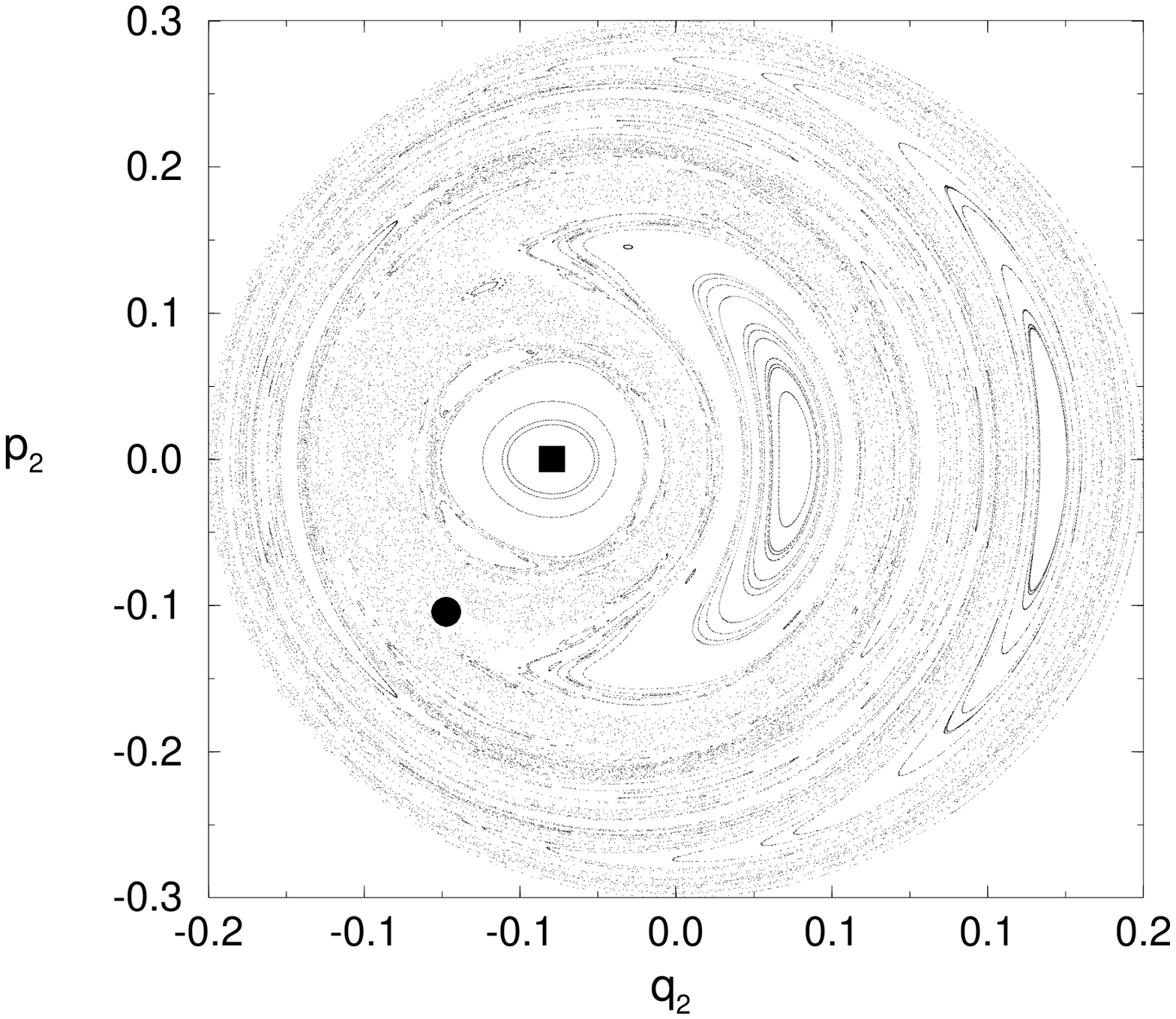}}
\vspace{0.3cm}
\caption{\small Poincar\'e section for the Nelson potential at $E=0.05$,
$q_1=0$ and $p_1>0$.
The symbols stand for the centers of the coherent states: filled
circle for chaotic region and filled square for regular region.}
\label{fig2}
\end{figure}

\begin{figure}[ht]
\begin{center}
\vspace{-0.3cm}
\includegraphics[scale=0.3,angle=-90]{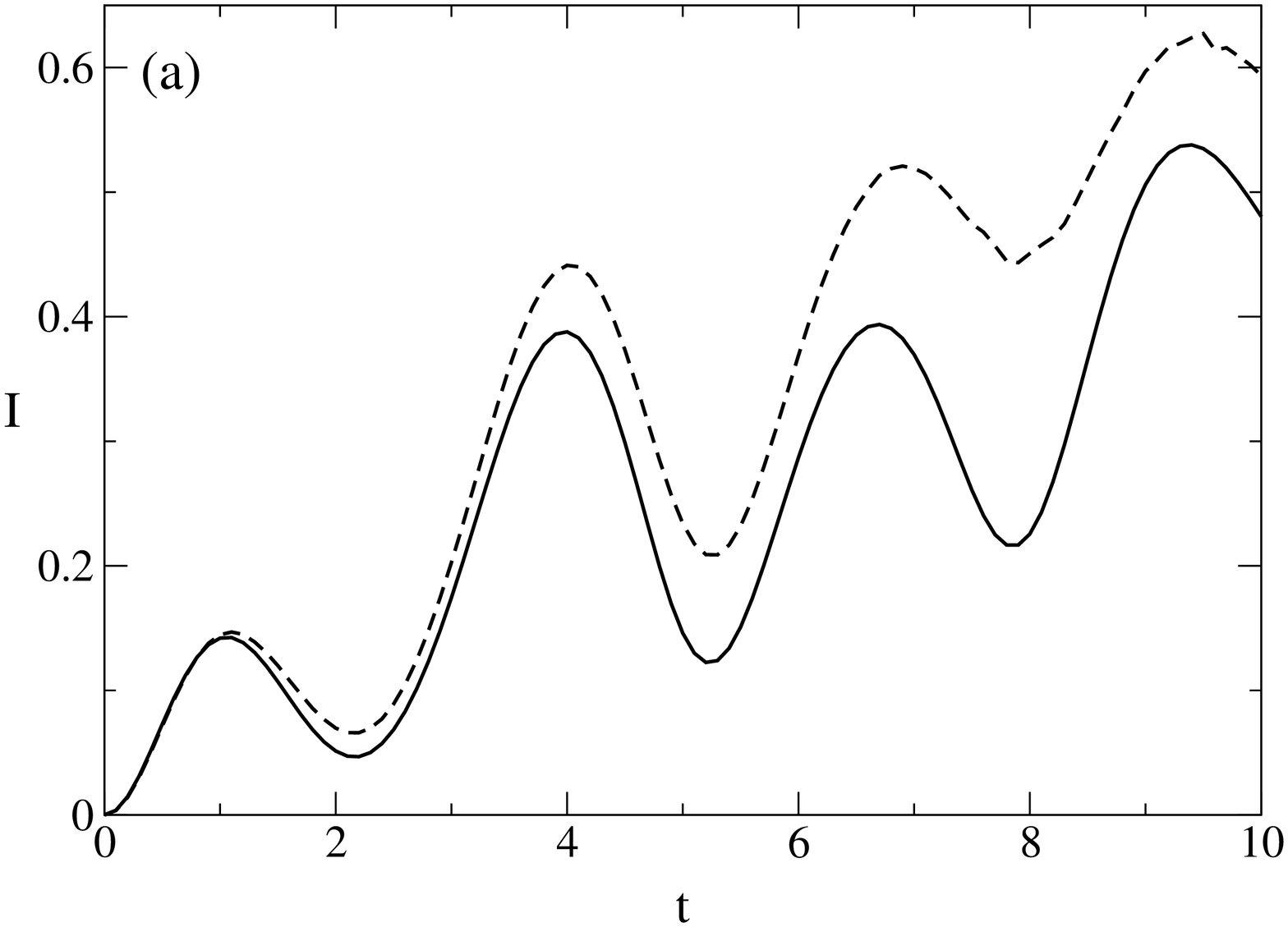} \\ \vspace{-0.3cm}
\includegraphics[scale=0.3,angle=-90]{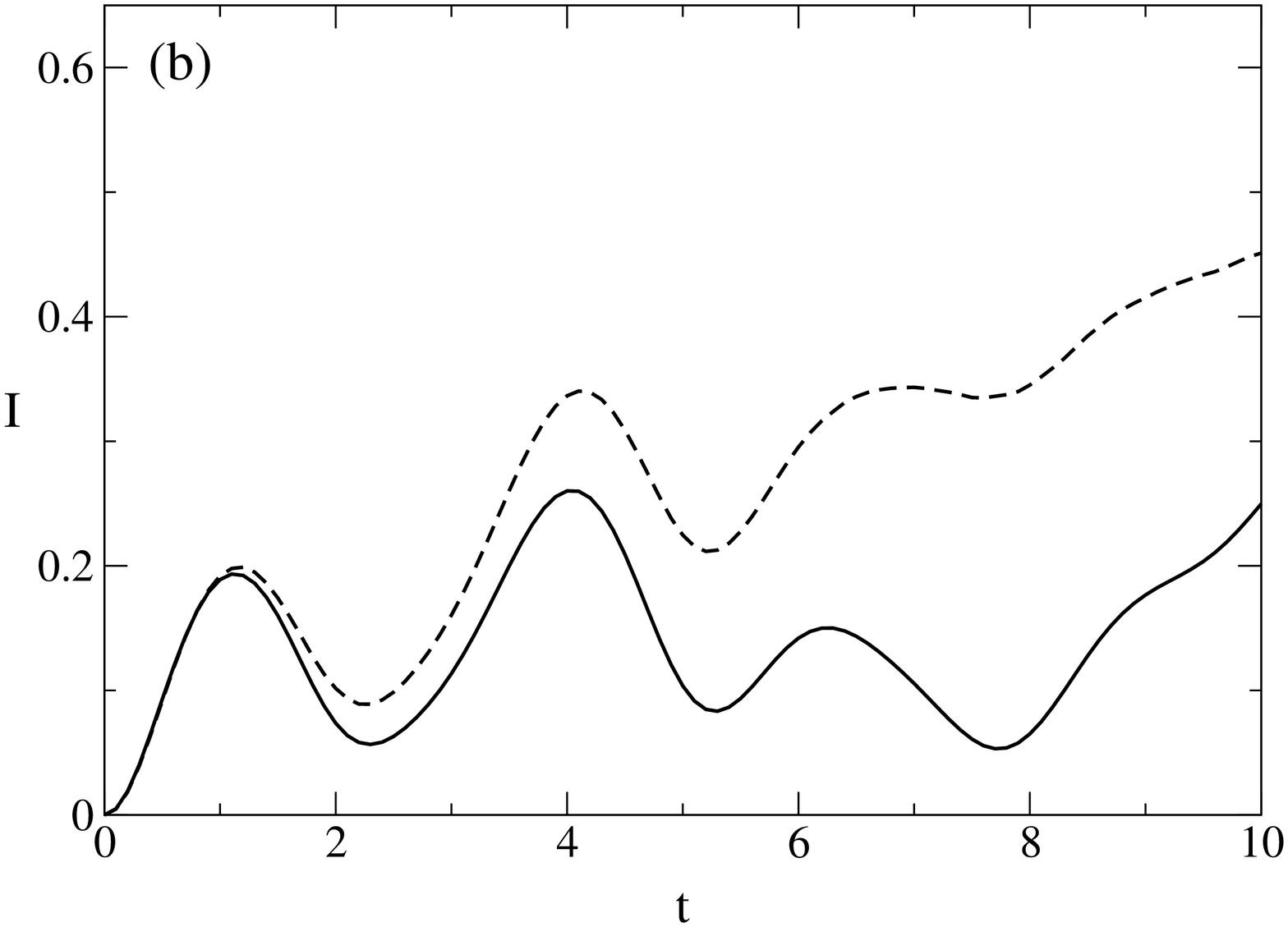}
\end{center}
\vspace{-0.3cm}
\caption{ QLMI (continuous line, $\hbar=0.05$) and CSLMI (dashed line)
for the nonlinear coupling and initial Gaussian distributions
centered on a chaotic orbit (a) and on a regular orbit (b).}
\label{fig3}
\end{figure}

\begin{figure}[ht]
\includegraphics[scale=0.3,angle=-90]{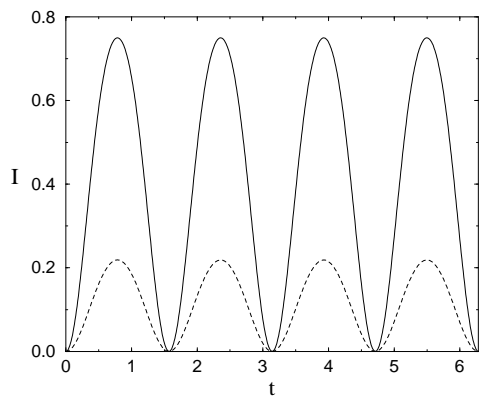}
\caption{\small Quantum and classical mutual linear information for the RWA
coupling for $w=\lambda=\hbar=1$ and initial state
$|\psi_0 \rangle=|0\rangle\otimes|1\rangle$.
The continuous line shows the quantum result $I(t)$ and the
dashed line the classical $I^{cl}(t)$.}
\label{fig4}
\end{figure}

\end{document}